\documentclass[onecolumn]{mn2e}  
\usepackage{epsf} 
\usepackage{amsmath} 
\usepackage{amssymb} 
\usepackage[first]{draftcopy} 
\begin{document}

\newcommand{\getinptfile}
{\typein[\inptfile]{Input file name}
\input{\inptfile}}

\newcommand{\mysummary}[2]{\noi {\bf SUMMARY}#1 \\ \noi \sl #2 \\ \capline 
	\hspace{-.13in} \raisebox{.0in}{$\sqcap$} \rm }  
\newcommand{\mycaption}[2]{\caption[#1]{\footnotesize #2}} 
\newcommand{\capline}{\mbox{}\hrulefill}
\newcommand{\mysection}[2]{ 
\section{\uppercase{\normalsize{\bf #1}}} \def\junksec{{#2}} } %
\newcommand{\mychapter}[2]{ \chapter{#1} \def\junkchap{{#2}}  
\def\thesection{\arabic{chapter}.\arabic{section}}
\def\thesubsection{\thesection.\arabic{subsection}}
\def\thesubsubsection{\thesubsection.\arabic{subsubsection}}
\def\theequation{\arabic{chapter}.\arabic{equation}}
\def\thefigure{\arabic{chapter}.\arabic{figure}}
\def\thetable{\arabic{chapter}.\arabic{table}}
}
\newcommand{\mysubsection}[2]{ \subsection{#1} \def\junksubsec{{#2}} }
\def\thenote{\addtocounter{footnote}{1}$^{\scriptstyle{\arabic{footnote}}}$ }

\newcommand{\myfm}[1]{\mbox{$#1$}}
\def\spose#1{\hbox to 0pt{#1\hss}}	
\def\ltabout{\mathrel{\spose{\lower 3pt\hbox{$\mathchar"218$}} 
     \raise 2.0pt\hbox{$\mathchar"13C$}}}
\def\gtabout{\mathrel{\spose{\lower 3pt\hbox{$\mathchar"218$}}
     \raise 2.0pt\hbox{$\mathchar"13E$}}}
\newcommand{\ltsim}{\raisebox{-0.5ex}{$\;\stackrel{<}{\scriptstyle \backslash}\;$}}
\newcommand{\simlt}{\ltsim}
\newcommand{\simgt}{\gtsim}
%
\newcommand{\unit}[1]{\ifmmode \:\mbox{\rm #1}\else \mbox{#1}\fi}
\newcommand{\ze}{\ifmmode \mbox{z=0}\else \mbox{$z=0$ }\fi }

%
\newcommand{\boldv}[1]{\ifmmode \mbox{\boldmath $ #1$} \else 
 \mbox{\boldmath $#1$} \fi}
%
\renewcommand{\sb}[1]{_{\rm #1}}%
\newcommand{\expec}[1]{\myfm{\left\langle #1 \right\rangle}}
\newcommand{\mone}{\myfm{^{-1}}}
\newcommand{\half}{\myfm{\frac{1}{2}}}
\newcommand{\nth}[1]{\myfm{#1^{\small th}}}
\newcommand{\ten}[1]{\myfm{\times 10^{#1}}}
\newcommand{\abs}[1]{\mid\!\! #1 \!\!\mid}
\newcommand{\as}{a_{\ast}}
\newcommand{\asr}{(a_{\ast}^{2}-R_{\ast}^{2})}
\newcommand{\bvm}{\bv{m}}
\newcommand{\calf}{{\cal F}}
\newcommand{\calI}{{\cal I}}
\newcommand{\calm}{{v/c}}
\newcommand{\calminf}{{(v/c)_{\infty}}}
\newcommand{\calQ}{{\cal Q}}
\newcommand{\calR}{{\cal R}}
\newcommand{\calw}{{\it W}}
\newcommand{\co}{c_{o}}
\newcommand{\cs}{C_{\sigma}}
\newcommand{\cst}{\tilde{C}_{\sigma}}
\newcommand{\cv}{C_{v}}
\def\dbar{{\mathchar '26\mkern-9mud}}	
\newcommand{\deldelr}{\frac{\partial}{\partial r}}
\newcommand{\deldelR}{\frac{\partial}{\partial R}}
\newcommand{\deldeltheta}{\frac{\partial}{\partial \theta} }
\newcommand{\deldelphi}{\frac{\partial}{\partial \phi} }
\newcommand{\ddotrc}{\ddot{R}_{c}}
\newcommand{\ddotxc}{\ddot{x}_{c}}
\newcommand{\dotrc}{\dot{R}_{c}}
\newcommand{\dotxc}{\dot{x}_{c}}
\newcommand{\Estar}{E_{\ast}}
\newcommand{\grpsi}{\Psi_{\ast}^{\prime}}
\newcommand{\kboltz}{k_{\beta}}
\newcommand{\levi}[1]{\epsilon_{#1}}
\newcommand{\limaso}[1]{$#1 ( a_{\ast}\rightarrow 0)\ $}
\newcommand{\limasinfty}[1]{$#1 ( a_{\ast}\rightarrow \infty)\ $}
\newcommand{\limrinfty}[1]{$#1 ( R\rightarrow \infty,t)\ $}
\newcommand{\limro}[1]{$#1 ( R\rightarrow 0,t)\ $}
\newcommand{\limrso}[1]{$#1 (R_{\ast}\rightarrow 0)\ $}
\newcommand{\limxo}[1]{$#1 ( x\rightarrow 0,t)\ $}
\newcommand{\limxso}[1]{$#1 (\xs\rightarrow 0)\ $}
\newcommand{\ls}{l_{\ast}}
\newcommand{\Ls}{L_{\ast}}
\newcommand{\mean}[1]{<#1>}
\newcommand{\ms}{m_{\ast}}
\newcommand{\Ms}{M_{\ast}}
\def\nb{{\sl N}-body }
\def\nbt{{\sf NBODY2} }
\def\nb1{{\sf NBODY1} }
\newcommand{\nuoned}{\nu\sb{1d}}
\newcommand{\ra}{\rightarrow}
\newcommand{\Ra}{\Rightarrow}
\newcommand{\rc}{r_{c} } 
\newcommand{\Rc}{R_{c} } 
\newcommand{\res}[1]{{\rm O}(#1)}
\newcommand{\rnsa}{(r^{2}-a^{2})}
\newcommand{\Rnsa}{(R^{2}-a^{2})}
\newcommand{\rs}{r_{\ast}}
\newcommand{\Rs}{R_{\ast}}
\newcommand{\Rsa}{(R_{\ast}^{2}-a_{\ast}^{2})}
\newcommand{\sa}{\sigma } 
\newcommand{\sac}{\sigma_{c} } 
\newcommand{\sas}{\sigma_{\ast} } 
\newcommand{\sasp}{\sigma^{\prime}_{\ast}}
\newcommand{\saxs}{\sigma_{\ast} } 
\newcommand{\sech}{{\rm sech}}
\newcommand{\tff}{t\sb{ff}} 
\newcommand{\ti}{\tilde}
\newcommand{\trel}{t\sb{rel}}
\newcommand{\ts}{\tilde{\sigma} } 
\newcommand{\tss}{\tilde{\sigma}_{\ast} } 
\newcommand{\vcol}{v\sb{col}}
\newcommand{\vs}{v_{\ast}  } 
\newcommand{\vsp}{v^{\prime}_{\ast}}
\newcommand{\vxs}{v_{\ast}  } 
\newcommand{\xs}{x_{\ast}}
\newcommand{\xc}{x_{c} } 
\newcommand{\xistar}{\xi_{\ast}}
\newcommand{\rmd}{\ifmmode \:\mbox{{\rm d}}\else \mbox{ d}\fi }
\newcommand{\rmD}{\ifmmode \:\mbox{{\rm D}}\else \mbox{ D}\fi }
\newcommand{\valfven}{v_{{\rm Alfv\acute{e}n}}}

%
\newcommand{\noi}{\noindent}
\newcommand{\bc}{boundary condition }
\newcommand{\bcs}{boundary conditions }
\newcommand{\Bcs}{Boundary conditions }
\newcommand{\lhs}{left-hand side }
\newcommand{\rhs}{right-hand side }
\newcommand{\wrt}{with respect to }
\newcommand{\iras}{{\sl IRAS }}
\newcommand{\cobe}{{\sl COBE }}
\newcommand{\Oh}{\myfm{\Omega h}}
%
\newcommand{\etal}{{\em et al.\/ }}
\newcommand{\eg}{{\em e.g.\/ }}
\newcommand{\etc}{{\em etc.\/ }}
\newcommand{\ie}{{\em i.e.\/ }}
\newcommand{\viz}{{\em viz.\/ }}
\newcommand{\cf}{{\em cf.\/ }}
\newcommand{\via}{{\em via\/ }}
\newcommand{\apriori}{{\em a priori\/ }}
\newcommand{\adhoc}{{\em ad hoc\/ }}
\newcommand{\viceversa}{{\em vice versa\/ }}
\newcommand{\versus}{{\em versus\/ }}
\newcommand{\qed}{{\em q.e.d. \/}}
\newcommand{\<}{\thinspace}
%
\newcommand{\km}{\unit{km}}
\newcommand{\kms}{\unit{km~s\mone}}
\newcommand{\kmsa}{\unit{km~s\mone~arcmin}}
\newcommand{\kpc}{\unit{kpc}}
\newcommand{\mpc}{\unit{Mpc}}
\newcommand{\hkpc}{\myfm{h\mone}\kpc}
\newcommand{\hmpc}{\myfm{h\mone}\mpc}
\newcommand{\parsec}{\unit{pc}}
\newcommand{\cm}{\unit{cm}}
\newcommand{\yr}{\unit{yr}}
\newcommand{\au}{\unit{A.U.}}
\newcommand{\AU}{\au}
\newcommand{\gm}{\unit{g}}
\newcommand{\solar}{\myfm{_\odot}}
\newcommand{\solarm}{\unit{M\olar}}
\newcommand{\Lsun}{\unit{L\solar}}
\newcommand{\Rsun}{\unit{R\solar}}
\newcommand{\seconds}{\unit{s}}
\newcommand{\micro}{\myfm{\mu}}
\newcommand{\micrometer}{\micro\mbox{\rm m}}
\newcommand{\Mdot}{\myfm{\dot M}}
%
%
%
\newcommand{\dgr}{\myfm{^\circ} }
\newcommand{\ddgr}{\mbox{\dgr\hskip-0.3em .}}
\newcommand{\mnt}{\mbox{\myfm{'}\hskip-0.3em .}}
\newcommand{\scnd}{\mbox{\myfm{''}\hskip-0.3em .}}
\newcommand{\hr}{\myfm{^{\rm h}}}
\newcommand{\dhr}{\mbox{\hr\hskip-0.3em .}}
%
%
%
%
%
%
%
\newcommand{\refindent}{\par\noindent\hangindent=0.5in\hangafter=1}
\newcommand{\figpar}{\par\noindent\hangindent=0.7in\hangafter=1}
%
%

\newcommand{\mybiblio}{\vspace{1cm}
		       \setcounter{subsection}{0}
		       \addtocounter{section}{1}
		       \def\junksec{References} 
 }

%
%
%

%
%
%
%
%

\newcommand{\vol}[2]{ {\bf#1}, #2}
\newcommand{\jour}[4]{#1. {\it #2\/}, {\bf#3}, #4}
\newcommand{\physrevd}[3]{\jour{#1}{Phys Rev D}{#2}{#3}}
\newcommand{\physrevlett}[3]{\jour{#1}{Phys Rev Lett}{#2}{#3}}
\newcommand{\aaa}[3]{\jour{#1}{A\&A}{#2}{#3}}
\newcommand{\aaarev}[3]{\jour{#1}{A\&A Review}{#2}{#3}}
\newcommand{\aaas}[3]{\jour{#1}{A\&A Supp.}{#2}{#3}}
\newcommand{\aj}[3]{\jour{#1}{AJ}{#2}{#3}}
\newcommand{\apj}[3]{\jour{#1}{ApJ}{#2}{#3}}
\newcommand{\apjl}[3]{\jour{#1}{ApJ Lett.}{#2}{#3}}
\newcommand{\apjs}[3]{\jour{#1}{ApJ Suppl.}{#2}{#3}}
\newcommand{\araa}[3]{\jour{#1}{ARAA}{#2}{#3}}
\newcommand{\mn}[3]{\jour{#1}{MNRAS}{#2}{#3}}
\newcommand{\mnras}{\mn}
\newcommand{\jgeo}[3]{\jour{#1}{Journal of Geophysical Research}{#2}{#3}}
\newcommand{\qjras}[3]{\jour{#1}{QJRAS}{#2}{#3}}
\newcommand{\nat}[3]{\jour{#1}{Nature}{#2}{#3}}
\newcommand{\pasa}[3]{\jour{#1}{PAS Australia}{#2}{#3}}
\newcommand{\pasj}[3]{\jour{#1}{PAS Japan}{#2}{#3}}
\newcommand{\pasp}[3]{\jour{#1}{PAS Pacific}{#2}{#3}}
\newcommand{\rmp}[3]{\jour{#1}{Rev. Mod. Phys.}{#2}{#3}}
\newcommand{\science}[3]{\jour{#1}{Science}{#2}{#3}}
\newcommand{\vistas}[3]{\jour{#1}{Vistas in Astronomy}{#2}{#3}}


\newcommand{\leftb}{<\!\!} \newcommand{\rightb}{\!\!>}

\newcommand{\oversim}[2]{\protect{\mbox{\lower0.5ex\vbox{%
  \baselineskip=0pt\lineskip=0.2ex
  \ialign{$\mathsurround=0pt #1\hfil##\hfil$\crcr#2\crcr\sim\crcr}}}}} 
\newcommand{\simgreat}{\mbox{$\,\mathrel{\mathpalette\oversim>}\,$}} 
\newcommand{\simless} {\mbox{$\,\mathrel{\mathpalette\oversim<}\,$}} 

\newcommand{\tcr}{t_{\rm cr}}

\newcommand{\erf}{{\rm erf}}

\newcommand{\sfe}{SFE} 

\newcommand{\df}{{\sc DF}}

\newcommand{\Fpk}{F_{\rm P/K}}

\def\emphasize#1{{\sl#1\/}}
\def\arg#1{{\it#1\/}}
\let\prog=\arg

\def\edcomment#1{\iffalse\marginpar{\raggedright\sl#1\/}\else\relax\fi}


\marginparwidth 1.25in
\marginparsep .125in
\marginparpush .25in

\reversemarginpar



%
\title{The impact of mass loss on star cluster formation. I. Analytic results } 
   \author[Boily \& Kroupa]{C.~M. Boily$^{1,\dagger}$ \&  P. Kroupa$^{2}$
\\          
$^{1}$Astronomisches Rechen-Institut, M\"onchhofstrasse 12-14 Heidelberg, D-69120 Germany 
\\
$^{2}$Institut f\"ur Theoretische Physik und Astrophysik der
Universit\"at Kiel, D-24098 Kiel, Germany 
\\ 
$^{\dagger}$Present address: Observatoire astronomique de Strasbourg, 11 rue de l'universit\'e, 67000 Strasbourg, France}

\maketitle

\begin{abstract} 
We study analytically the disruptive effect of
instantaneous gas removal from a cluster containing O stars. 
 We setup an iterative calculation based on the stellar velocity distribution function to
compute the fraction of stars that remain bound once the cluster has
ejected the gas and is out of equilibrium. We show that the stellar 
bound fraction is a function of the initial cluster distribution function as well 
as the star formation efficiency, $\epsilon$, taken constant throughout the cluster. 
 The case of the Plummer sphere is dealt with in greater details. We find 
for this case that up to $\simeq 50\%$ of the stars  may remain bound when $\epsilon$ 
assumes values $< \half$, contrary to expectations derived from the virial 
theorem. The fraction of bound stars is expressed algebraically for polytropic 
distribution functions.\end{abstract} 

\section{Introduction}
Most if not all stars are formed in clusters or associations, 
 and therefore this is the predominant 
mode of star formation contributing to the Galactic-field population (see reviews in Grebel \& Brandner 2002).
To understand  how such aggregates form and evolve remains a
severe challenge however, since no  theory exists yet 
that accounts in detail for the (presumably) simpler case of the formation of
individual stars. On the observational front, surveys of star-forming
regions suggest that the mass-fraction of gas used up at birth does
not exceed 30\% of the total when averaged over typical cluster scales (Lada 1999).  This low
star-formation efficiency (\sfe) implies that   young stars or clusters
of stars should be embedded in gas (see Lada \& Lada 1991). 
  Yet populous clusters with ages $\simless$ a few Myrs
(e.g. the Orion Nebula Cluster, R136 in 30~Doradus) are already void
of gas. The mass-fraction of gas left after the epoch of star formation 
 depends crucially on the star formation history of the cluster. 
The formation of proto-stars by hydrodynamical collapse 
points to the rapid formation of opaque stellar cores followed by time-dependent 
accretion, with the more massive cores  accreting at higher  rates 
 (e.g. Boily \& Lynden-Bell 1995). 
Stellar masses are then accrued over 
a period of typically $10^5 - 10^6$ years (Low \& Lynden-Bell 1976;  
 Shu 1977; Wuchterl \& Klessen 2001). 
 Observational evidence in support of rapid accretion as opposed to a slow star formation 
process sustained by ambipolar diffusion of magnetic field is discussed by Hartmann et al. (2001), 
who point out that the age spreads between cluster members are too small compared with 
that expected from ambipolar diffusion (see Jones et al. 2001 and references therein).  
Because opaque stellar cores will not all have the 
same mass (or indeed chemical composition), it is unlikely that  stars will 
 end their accretion phase at the same universal time. Consequently, 
 the star cluster will  not be entirely depleted of gas 
when the more massive stars reach the Hydrogen-burning phase. \newline 

 OB stars provide an efficient mechanism by which gas is driven out. 
Stellar winds from  OB stars yield a momentum flux ${\cal F} \sim 8\times 10^{-3} M_{\sun}\,
{\rm km\, sec}^{-1} {\rm yr}^{-1}$ (Churchwell 1999). This is
sufficient to unbound gas from an $10^3\,M_\odot$, 1 pc radius
 cluster in $\approx 10^5$ years, which is comparable to its
crossing time $\tcr = 2 R/\sigma \propto 1/\sqrt{G\rho(0)}$. Here $\rho(r)$ is the mass density 
and $\sigma$ the velocity dispersion.  In addition, the ionising radiation heats the
gas to $10^4$~K, causing an over-pressure and expansion at the
sound velocity $\approx 10$~km/s.  
  Furthermore, the star-formation timescale  of $\sim 10^6$ yrs far exceeds  
a single cluster crossing time $\tcr$: the magnitude of stellar velocities should therefore reflect the 
depth of the cluster gravitational potential since protostars would have time to explore the entire cluster volume. This suggests that  a rich 
 cluster is already close to virial equilibrium when OB winds set in.  \newline 

 The question whether or not the stars will re-establish a  bound gas-free  equilibrium 
has  bearings on the  spatial distribution and statistics  of  stellar populations 
in the Galaxy (Kroupa 2002,  Grebel \& Brandner 2002). A long-standing goal has been 
to understand  in what way gas-ejection and  the low \sfe\ inferred from observations set 
constraints  on  the star cluster properties at birth  (see e.g. Lada 1999).    
Hills (1980) has argued from this standpoint and using 
 the virial theorem that if the \sfe\ is less or equal to 50\%  the entire 
 cluster will dissolve. His classic argument holds for clusters that have 
 achieved virial equilibrium and which then undergo a sudden loss of mass. Theoretical and numerical 
work support 
this conclusion when gas evacuation proceeds rapidly 
(Tenorio-Tagle et al. 1986; Geyer \& Burkert 2001),  while the numerical N-body study of Lada et al. (1984) already 
found bound stellar cores regardless of the gas evacuation time-scales. 
  This was also found to be the case in  recent, larger-N studies (Goodwin 1997; Kroupa, Aarseth \& Hurley 2001).   Adams (2000) has recently pointed out that if the \sfe\ peaks 
 with the central cluster density, a small core of stars will always remain 
 bound, even for an \sfe\ well below the reference 50\% obtained from the virial. 
 This effectively solves the problem of forming bound clusters but at the expense of 
simplicity, by assuming a radial variation for the \sfe, the nature of which is yet poorly 
constrained. Our goal is to show that one may solve for the bound fraction of stars analytically 
without  introducing further complications. \newline 
   
 The numerical N-body results and the analysis of  Adams (2000) 
   hint that Hills' (1980) original argument is lacking an 
ingredient to account for bound clusters at low \sfe. We note that the derivation 
  based on the virial theorem 
 is a statement about the global properties of the cluster with regards to the impact of the  
\sfe. A refinement is therefore possible if 
 we reformulate the problem from the point of view of the equilibrium 
stellar distribution function $(\equiv$\df), $F(\boldv{x},\boldv{v})$, as follows~: 
 First, select stars by binding energy $(E)$ to identify bound  ($E<0$)
 and candidate escaping  ($E>0$) stars. The \df\ may  then be modified  
 to take account of the cluster mass loss due to unbound gas and stars. 
 As a second step, 
    we setup an iterative scheme to identify bound stars self-consistently once gas and  escapers 
are subtracted from the gravitational potential. This scheme allows us to  
 solve analytically for the surviving fraction of bound stars. We show that 
the result is function of both \sfe\ and the chosen \df\ by contrasting results obtained 
for polytropes, and in particular the Plummer sphere. The time-evolution 
of clusters, and applications to other \df's, are deferred to a companion paper (Paper II). 

\section{Hills' problem} 
A cluster of stars forms converting a fraction $\epsilon$ of the gas
into stars in the process. Stellar winds or a supernova event blow/s
out the remainder on a time-scale $\tau \ll \tcr$. What fraction of the
stars remain to form a bound cluster? Since the star formation epoch
will have lasted as long as or longer than a cluster dynamical time
$\tcr$, the system as a whole is close to virial equilibrium
before gas is expelled. In the following,  we 
consider the limit $\tau \rightarrow 0$, which defines an initial value 
 problem for an open cluster's mass profile and velocity field. \newline 

Hills (1980) argued that equilibrium, self-gravitating stellar
systems would  dissolve if more than half the mass is lost instantaneously: 
 stars then preserve their kinetic energy established under a 
deeper potential well, hence may escape if their binding energy $E$ 
becomes positive. We write the total energy  

\begin{equation}
 E = -  \frac{GM^2}{R} + \frac{1}{2} M\langle v^2\rangle < 0\ ,\label{eq:energy}
\end{equation} 
where $M, R$ are the mass and gravitational radius of a spherical
distribution of mean square velocity $\langle v^2\rangle$.
  Before gas-expulsion the total mass in gas and stars is
$M  = M_{\rm init} =  
M_{\rm gas}+M_\star$, while after the gas is expelled
$M=M_\star$ but the velocity dispersion, $\langle v^2\rangle = 
GM_{\rm init}/R$, remains unchanged.  The new, total system 
energy is now 

\begin{equation} E_\star = - \frac{G M_\star^2}{R} + \frac{1}{2} M_\star \langle v^2 \rangle \equiv - \frac{GM_\star^2}{2 R_\star} \label{eq:energy_s} \end{equation} 
where the last equality follows from applying the virial theorem to the 
 stellar system of mass $M_\star$ after it has reached a new equilibrium, of energy $E_\star$ but with radius $R_\star > R$. We may rearrange (\ref{eq:energy_s}) to obtain 

\begin{equation} \frac{R_\star}{R} = \frac{1}{2} \frac{M_\star}{M_\star - \half M} = \frac{1}{2}\frac{M- M_{\rm gas} }{\half M - M_{\rm gas}} \ . \label{eq:ratio} \end{equation} 
The system radius $R_\star \rightarrow \infty $ when $M_{\rm gas} = M/2$ and hence 
the stellar system has zero or positive total binding energy 
when  the 
mass is reduced by 50\% or more $(\epsilon = M_\star/M \le 1/2)$. 

Hills' derivation applies  to any mass distribution. 
It does not, however, say anything specific about the evolution of the mass profile except 
to relate the final virial radius $R_\star$ to the initial system radius $R$. 
Our intuition is that a fraction of the stars with larger velocities will 
 escape before we reach the reference 50\% gas mass loss and hence the residual stellar potential 
should  not be evaluated from the global stellar mass fraction alone. In other words, the 
stellar velocity \df\ should determine  which stellar orbit remains bound 
according to its binding energy: this 
 is confirmed when considering orbits in a harmonic potential as shown in Appendix A. 
 The example of harmonic motion 
 will serve us later in interpreting results obtained for systems with realistic 
mass profiles and \df's. Below we introduce a scheme to pin down more accurately the 
relation of bound mass to star formation efficiency, $\epsilon$. 

\section{Distribution functions} \label{sec:df}  

\subsection{Iterative scheme} 
Consider an isotropic mixture of stars and gas within a spherical volume. The stellar \df\ $F(\boldv{r},\boldv{v})$ is then a function of energy alone. 
As a simplifying assumption we take the \sfe\ to be independent of
position in the cluster. Hence the \sfe\ parameter $\epsilon = $ constant. 
Thus gas and 
stars are initially mixed in the same proportion throughout, however
note that $\epsilon$ does not fix the profiling of stellar masses with
radius, rather the total mass of gas made into stars~:

\[ \frac{\sum_{\rm i} ({\rm 
number\ of\ stars\ in\ mass\ bin\ } i) \times ({\rm stellar\ mass\ }\
m_i)}{m_{\rm gas}} = {\rm constant } \] 
at any radius $r$. Both gas and stars are distributed according to the
same \df. The total system mass is then

\begin{equation} M = M_\star + M_{\rm gas} = \epsilon^{-1}\, M_\star = 
 \epsilon^{-1}\ \iint_V F(\boldv{r},\boldv{v}) \rmd^3\boldv{r}\rmd^3\boldv{v} = \epsilon^{-1}\ 
\int_{E_c}^{0}
F(E)\, {\cal W}(E)\, {\rm d} E, \label{eq:defepsilon} \end{equation} 
where $|E_c|$ is the maximum binding energy and ${\cal W}$ the density of states of energy $E$. The cluster potential $\phi(r)$ follows from integrating twice Poisson's equation

\begin{equation} 
\phi(r ; M, R ) = \int_\infty^{r}\frac{{\rm d}u}{u^2} \int_0^u {4\pi
G\rho(w)w^2{\rm d}w} \equiv y(r/R)\, \frac{G M}{R}, \label{eq:phi}
\end{equation}
where $y(x)$ is a dimensionless function. Keeping $R$ constant while 
removing the gas instantly so the potential includes stars only, we have 

\begin{equation}
\phi(r; M, R ) \rightarrow \phi_\star = \epsilon\, \phi, \label{eq:phistar}
\end{equation}
and hence the fraction of stars at radius $r$ which now have positive
energy is

\begin{equation} 
1 - \lambda_e = \frac{\displaystyle \int_{v_{e,\star}}^{v_e} F(\boldv{r},\boldv{v}) \,  \rmd^3\boldv{r}\, v^2\,{\rm d}v }{ \displaystyle \int_{0}^{v_e} F(\boldv{r},\boldv{v})\,  \rmd^3\boldv{r}\, v^2\, {\rm d}v} = \frac{\displaystyle \int_{v_{e,\star}}^{v_e} f(v) \, v^2\,{\rm d}v }{ \displaystyle \int_{0}^{v_e} f(v)\, v^2\, {\rm d}v}\, ,
\label{eq:escape} \end{equation}
with $v_{e}$ the (local) escape velocity computed from $\phi$ (before
gas-expulsion), and similarly for $v_{e,\star}$ obtained from 
$\phi_\star$. For convenience of notation, we write the stellar velocity distribution function $f(v)$ defined 
at constant radius $\boldv{r}$ as 

\[ f(\boldv{v}) \rmd^3\,\boldv{v} \equiv F(\boldv{r},\boldv{v}) \rmd^3\boldv{r}\,\rmd^3\boldv{v} \ . \] 

For a given \df\ and $\epsilon$, we may compute the quantity $1 -
\lambda_e$ at all radii. Note that should the \sfe\ be a function of
radius, the re-normalisation (\ref{eq:phistar}) leading to $\phi_\star$ would
not apply, however (\ref{eq:escape}) may still be computed if the
stars' potential is given and $v_e$ known from the initial gas + stars
mixture.  The key step is to adjust the potential $\phi_\star$ itself,
since potential and velocity field do not match any longer. This is
normally computed numerically as an initial value problem, 
for instance using N-body integration which takes into account all 
 star-star interactions and the redistribution of energy between 
them in the time-dependent potential. We take a different approach. 
 If we thought that stars acquire
only little extra energy during the time that they escape, then the fraction
of stars remaining might be computed as follows. 

Since the fraction (\ref{eq:escape}) of positive-energy stars is known
at all $r$, we re-compute the gravitational potential counting only 
stars with $E \le 0$ at each radius. Neglecting dynamical evolution, 
the cluster
radius $R$ and structure function $y(r/R)$ are
 unchanged and hence the potential can be re-computed by
integration, from $\infty$, inwards. Once the new potential is known, a
fraction of the remaining stars will again be unbound by virtue of the
stars lost during the previous iteration. 
We  therefore repeat the procedure until
finally the cluster mass converges to a finite quantity, in which case
no more stars escape and the original distribution function is
depleted from all escapers in a self-consistent manner.

We consider the  case of scale-free \df\ in detail, then turn to 
 more physically-motivated cases in the following sections. 

\subsection{Example : scale free \df's} 

The case of a power-law  isotropic velocity distribution, 
\[ f(\boldv{v})\,\rmd^3\boldv{v} \propto v^{\beta-2} \rmd^3\bold{v} \propto v^{\beta}\rmd{v}\, ,\]
 where $\beta$ is constant,
provides a tractable starting point. The \df\ is 
truncated at the local escape velocity and we take the potential $\phi(r)$ to be a power of 
the radius: the system is therefore scale-free. The one-dimensional velocity dispersion
obtained from Jeans' equation which for power-law density and potential  yields 

\begin{equation} \sigma^2(r) =  \frac{1}{\rho(r)} \, \int_r^\infty \rho(x) \nabla\phi(x) \rmd x = - 2 K\, \phi(r) = K\, 
v_e^2 \label{eq:jeans} \end{equation}
 at each radius for some constant of proportionality, $K$.
Substituting $\rho(r) \rightarrow \epsilon \rho(r)$ in (\ref{eq:phi}), we find the new potential
from (\ref{eq:phistar}) and escape velocity            

\begin{equation} v_{e,\star}(r) = \sqrt{2
\phi_\star(r)} = \epsilon^{1/2} v_e(r)\, . \label{eq:scaleve} \end{equation} 
 Equation (\ref{eq:escape})  becomes (with $u \equiv v/v_e$)

\begin{equation} 
1 - \lambda_e = \frac{\displaystyle \int_{v_{e,\star}}^{v_e} f(v) \,
v^2\,{\rm d}v }{ \displaystyle \int_{0}^{v_e} f(v)\, v^2\, {\rm
d}v} = \frac{\displaystyle \int_{\epsilon^{1/2}}^{1} f(u) \, u^2\,{\rm
d}u }{ \displaystyle \int_{0}^{1} f(u)\, u^2\, {\rm d}u} = 1 -
\epsilon^{(\beta+1)/2},
\label{eq:escapepower} 
\end{equation}
where we took $\beta > -1$. Thus $\lambda_e = \epsilon^{(\beta+1)/2}$ is 
independent of radius. Repeating the procedure to take account of the
positive-energy stars, we substitute $\epsilon \rightarrow \lambda_e \cdot
\epsilon$, etc, so that after $j$ iterations 

\[ \lambda_e[j] = (\lambda_e[j-1]\,\epsilon)^{(\beta+1)/2}\,  \] 
with $\lambda_e[0] = 1$. Assuming convergence of the series with $j\rightarrow\infty$ we may solve 
for $\lambda_e$ to obtain 

\begin{equation} \lambda_e = \epsilon^{\displaystyle\frac{1+\beta}{1-\beta}}\, . \label{eq:powersol}  \end{equation}
(See Boily \& Kroupa 2002 for an alternative derivation.) 
 In terms of the stellar mass fraction, the predicted mass of bound stars, $M^b_\star$, is then simply 

\begin{equation} M^b_\star = \lambda_e\, \epsilon\, M = \lambda_e\, M_\star 
\label{eq:powerlaw}
\end{equation} 
(and similarly for the stellar density $\rho_\star[r]$ wrt $\rho[r]$).
The eigenvalue $\lambda_e$ in (\ref{eq:powerlaw}) converges to a   
non-zero (positive) value as $j\rightarrow\infty$ only for  $\beta <
1$, since $\epsilon \le 1$. All \df's with $\beta > 1$ lead to cluster
disruption, because the high-velocity range of the \df\ is too densely
populated, leading to catastrophic stellar loss after the expulsion of
any amount of gas. 

Fig.~1 graphs the fraction of 
bound stars $M^b_\star/M_\star$ obtained from (\ref{eq:powerlaw}) as 
function of the \sfe\ $\epsilon$ for five values of the power index $\beta$. All curves meet at $\epsilon = 1$  and 
$M^b_\star/M_\star = 1$. Note how a flat \df\ ($\beta = 0)$ predicts 
a linear decline of bound star fraction. 

The volume density  $\rho(r)$ is recovered from integrating 
 $F(\boldv{r},\boldv{v})$ over all velocities  at constant radius $r$ : 

\begin{equation} \rho(r) \rmd^3\boldv{r} =  \int_0^{v_e} F(\boldv{r},\boldv{v})\,\rmd^3\boldv{r}\,\rmd^3\boldv{v} =  \int_0^{v_e} f(v)\, 4\pi\, v^2\rmd v \propto \int_0^{v_e}  v^\beta \rmd v =  \frac{v_e^{\beta+1}}{\beta+1} =  (-2\phi[r])^{(\beta+1)/2} \equiv (-2\phi)^n \label{eq:pseudopoly} \end{equation} 
 where the last relation emphasises the similarity with polytropes of 
index $n$, to which we return. 
 Since (\ref{eq:powerlaw}) was derived  assuming a power-law potential, it follows from (\ref{eq:pseudopoly}) 
 that $\rho(r)$ is also a power of
the radius. Power-law density profiles may be of application to young 
clusters such as NGC2282 (Horner et al. 1997).  Letting $\rho(r) \propto r^{-\alpha}$, we deduce from (\ref{eq:phi}) that $\phi(r) \propto r^{2-\alpha}$; a centrally condensed cluster requires $\alpha > 0$. 
 The power indices $\alpha$ and $\beta$ are then related to each other 
 using (\ref{eq:pseudopoly}) 

\[ \alpha = 2\, \frac{\beta+1}{\beta-1}\,  \] 
and hence $\alpha > 0$ for $\beta > 1 $ or $\beta < -1 $. We recall that $\beta > -1$ from (\ref{eq:escapepower}), hence 
 we must choose $\beta > 1$ for a self-consistent solution. NGC2282 shows an  
 axially-averaged 
 surface density $\Sigma\propto R^{-1}$ which can be obtained from projecting a
spherical cluster of volume density 
  $\rho \propto r^{-2}$, or $\alpha = 2$ (singular isothermal sphere).  
The index 
$ \beta = (\alpha + 2)/(\alpha -2) \rightarrow \infty$ when $\alpha = 2$ which is unrealistic. 
  Therefore no  self-consistent solution of this sort exists with $\beta > -1$ 
  applicable to NGC2282.  The large value of $\beta$ obtained for this case 
 is at best indicative of rapid dissolution for any $\epsilon < 1$. \newline 

The example of power-law \df's suggests a strong dependence of the effective \sfe\ when cluster 
disruption occurs, and the shape of the \df\ itself.  
  Naive application to NGC2282 of a power-law \df\ 
based on the Jeans 
equations of equilibria leads to a nonphysical solution.   In the following we 
 apply our iterative scheme to other known \df's. 

\section{Plummer model and the family of polytropes} 
We wish to extend  our basic result (\ref{eq:powersol}) to a 
class of \df's with properties similar to those of observed star 
clusters. The power-law relation  of the scale-free \df's suggests 
that we consider the family of polytropes defined as 

\begin{equation} F(E) = (-E)^{n-\frac{3}{2}} \label{eq:polytrope} 
\end{equation} 
where the power index $n > 1/2$ may take non-integer values (BT+87). 
The case $n = 5$ has been studied in detail by Plummer (1911). Then 

\begin{equation} 
 F(E) = (-E)^{7/2} = ( |\phi(r)| - \half\,v^2)^{7/2}\, .\label{eq:plummer} \end{equation} 
 The total system mass of a Plummer sphere is finite but infinite in extent; 
   the density profile 

\begin{equation} \rho(r) = \frac{3M}{4\pi R_p^3} \, \left[ 1 + r^2/R_p^2 \right]^{-5/2} \propto \phi^5(r) = \left( \langle 2 v^2[r]\rangle \right)^5 \label{eq:plummer_rho} \end{equation} 
where $M$ is the total system mass, $R_p$ the Plummer radius  and both 
the velocity dispersion and density maximise at the centre  (see e.g. Spitzer 1987).   
At large distances, the density profile $\rho(r) \propto r^{-5}$ decreases like a power of the 
radius. Remembering (\ref{eq:jeans}), 
the Plummer velocity \df\ $f(v) \propto \ ( 1 - [v/v_e]^2)^{7/2}$. 
Inserting this into (\ref{eq:escape}) and with  

\[ p(\epsilon) \equiv  1 -\frac{1210}{105}\, \epsilon
+ \frac{2104}{105}\, \epsilon^2 - \frac{1488}{105}\, \epsilon^3 + \frac{384}{105}\, \epsilon^4\]
 we find
\begin{equation} 
1 - \lambda_e = \frac{\displaystyle \int_{\epsilon^{1/2}}^{1} ( 1 - u^2
)^{7/2} \, u^2\,{\rm d}u }{ \displaystyle \int_{0}^{1}( 1 - u^2
)^{7/2} \, u^2\, {\rm d}u} = 1 - \frac{2}{\pi} \left( { \sin^{-1} 
\epsilon^{1/2} - p(\epsilon)\,\epsilon^{1/2}\sqrt{1-\epsilon} }{}\right) . \label{eq:plummer_lambda}
\end{equation} 
We note that, as for the  case of power-laws, 
the solution is independent of radius. 
Therefore the potential
$\phi_\star$ and  escape velocity $v_e$ follow from 
(\ref{eq:phistar}). 
 The fraction of stars labeled for escape are removed from the potential and the integral 
re-evaluated with the substitution of $\epsilon^{1/2} $ by $ (\epsilon\, \lambda_e)^{1/2}$ as the lower 
bound of integration in (\ref{eq:plummer_lambda}).  This
 follows from the same transformation leading to 
(\ref{eq:scaleve}). 
We may repeat the procedure until $\delta(1 - \lambda_e) \rightarrow 0$ on successive iterations so no additional stars are lost.  The converged fraction $\lambda_e$ is then solution of 

\begin{equation} 
 \lambda_e =  \frac{2}{\pi} \left( { \sin^{-1} 
(\lambda_e\epsilon)^{1/2} - p(\lambda_e\epsilon)\,(\lambda_e\epsilon)^{1/2}\sqrt{1-\lambda_e\epsilon} }{}\right) . \label{eq:lpara} \end{equation} 
This may be expressed in parameterised form. Defining 

\begin{equation}
 x \equiv \lambda_e\, \epsilon\ ; \ x \subseteq [0, 1]\, , \label{eq:defx}
\end{equation} 
we have  the solution $\lambda_e(x)$  from (\ref{eq:lpara}). 
 The full solution follows from solving  for $\epsilon$ as a function of 
$x$, 

\begin{equation} \epsilon(x) = \frac{x}{\lambda_e(x)} \label{eq:epara}
\end{equation} 
which is known 
for all $x$ in the allowed range. The relation $\lambda_e(x)$ to $\epsilon(x,\lambda_e)$ 
saves us from iterating on (\ref{eq:lpara}), which otherwise yields the same result. 
 In Appendix B we extend the solution (\ref{eq:lpara}) 
to configurations out of dynamical equilibrium. 


\subsection{Analytic results for Plummer spheres} 
From the  definition (\ref{eq:defx}) we must have
 $\epsilon = \lambda_e = 1 $ when $x = 1$.  
Inserting (\ref{eq:lpara}) into (\ref{eq:epara}) we find 

\[ \lim_{x\rightarrow 0} \epsilon(x) \propto x^{-1/2} \rightarrow \infty \]
while $\lambda_e(x\rightarrow 0) = 0$. Since $\epsilon(1) = 1$ and $\epsilon(x<1) \le 1$ for 
a physically meaningful solution, it must reach 
a minimum somewhere in the interval $x: [0, 1]$. We find $\rmd\,\epsilon/\rmd x \equiv \epsilon^\prime(x) = 0$ from basic calculus at   

\[ x = 0.225259 \ ; {\rm minimum\ of\ }  \epsilon = 0.442201 \] 
when $\lambda_e = 0.509404$. Solutions exist only for 
 an \sfe\ in the range $\epsilon: [0.442201, 1]$. 
The result (\ref{eq:epara}) is displayed on Fig.~2 where we graph the same quantity $\lambda_e$ as in Fig.~1 but for a Plummer \df. The curves are best 
understood if one starts at $(\epsilon,\lambda_e) = (1,1)$, upper right-hand corner of the plot, shifting to the left as the \sfe\ $\epsilon$ is reduced. 
The most striking feature is the sudden drop 
in bound fraction $\lambda_e$ 
for an \sfe\ $\approx 44.22$\% (solid curve, Fig.~2). 
 The lower branch of the parameter solution $\lambda_e(x)$ is shown 
as dash but corresponds to no physical state of the system. For comparison, we have also displayed as open 
circle the results of a series of N-body calculations, which show remarkable agreement with (\ref{eq:lpara}). 
Details of the N-body calculations will be found in Paper II. 

Around the critical value, the predicted bound fraction varies from 
$\lambda_e \approx 0.63$ upward of an \sfe\ $\epsilon = 0.443$, to 0 for an \sfe\ $< 0.442$. For $\epsilon = 0.50$, we  compute a 
fraction of bound stars $\lambda_e \approx 86\%$. 
This counts as bound nearly all stars initially in 
 the system, when we expected complete dissolution 
from (\ref{eq:ratio}). \newline 

To illustrate the significance of repeated iterations on (\ref{eq:plummer_lambda}), we also plot on the figure the result when no iterations are performed and the 
solution $\lambda_e(\epsilon)$ taken directly from (\ref{eq:plummer_lambda}) (dotted line, Fig.~2). This graphs a selection of stars by energy but without 
taking into account the  shallower  gravitational potential as a result of 
stars leaving the system. As the bound fraction  
remains finite for any positive \sfe, and 
no sudden dissolution is predicted, we conclude that feedback  on the gravitational potential from  stars
 lost  changes the character of the solution very significantly. This should not come 
as a surprise since cluster dissolution implies a stream of stars moving outwards. 
 What this shows is that the impact of this stream must be taken into account when 
 quantifying  the fraction of stars which ultimately  
 will form a bound equilibrium after a phase of dynamical evolution. By contrast, Adams (2000) allows $\epsilon$ 
  to vary with position but does not iterate to establish final membership. 

\subsection{Analytic results for all polytropes}
 The results of the previous section may be extended to all polytropes of index 
$n > 1/2$ defined by (\ref{eq:polytrope}). The general solution $\lambda_e(x)$ 
may be written for any index $n$ according to the definition (\ref{eq:defx}). 
 We find 

\begin{equation} \lambda_e(x) = \frac{ x^{3/2}\, _2F_1 \left( \frac{3}{2}, \frac{3}{2}-n, \frac{5}{2}, x \right) }{\, _2F_1 \left( \frac{3}{2}, \frac{3}{2}-n, \frac{5}{2}, 1 \right)} 
 = \frac{\Gamma(n+1)}{\Gamma(n-\half)}\, \frac{4\, x^{3/2}}{3\sqrt{\pi}}\, _2F_1 \left( \frac{3}{2}, \frac{3}{2}-n, \frac{5}{2}, x \right)
\end{equation} 
where $\Gamma(u)$ is the $\Gamma-$function, and the hyper-geometric function 
$\, _2F_1(k,l,m,z)$ assumes  polynomial expressions for integer values of $n$ (see e.g. Weisstein 1999 for details). 
 We may extend the steps for the Plummer sphere to any polytrope, 
 identifying the minimum of $\epsilon(x)$ in each case. This is done easily with 
software such as {\sc mathematica} or {\sc maple}. 

 Results for $x$ where $\epsilon^\prime = 0 $ are given in Table~\ref{tab:polytropes} 
for several polytropes of index $n$ ranging from $n = \frac{3}{2}$ to $40$. 
 The minimum of $\epsilon$  is also listed along with 
the stellar mass fraction $\lambda_e$. The case of $n = 3/2$ is special. Then 
 $\epsilon =  x^{-\half}$ everywhere in the allowed interval, and so $\epsilon^\prime (x) < 0 $ 
 for all $x$'s. Indeed no solutions exist with $\epsilon(x) < 1$ for all indices $n$ in the range $ \half < n \le \frac{3}{2}$. The $n = \frac{3}{2}$ polytrope admits $\epsilon = \lambda_e = 1$ as a special solution satisfying our constraints on bound mass, however $\epsilon^\prime(1) \ne 0 $ does not satisfy the mathematical requirement of a local minimum. 

All polytropes of index $n \ge 5 $ stretch to infinity, and those with $n > 5$ all 
have infinite mass.  
 We note that $\epsilon$ decreases smoothly with increasing $n$ and 
 in particular $\epsilon \rightarrow 0$ as $1/n$. The same is true of the mass fraction $\lambda_e$, though 
for $n = 40$ we find $\lambda_e \approx 0.40$ for a relatively low \sfe\ of $\epsilon \approx 5\%$. 
 The bound fraction $\lambda_e$ at low \sfe\ recovered from the polytropic \df\ 
overlaps with observational estimates if the index $n$ is sufficiently large. However, 
all polytropes with $n < 5$ dissolve at a larger \sfe\ than the Plummer 
case ($n = 5$). 
  Since $n<5$ polytropes show a  broad region of near-uniform volume density, these data 
suggest that the core-halo structure plays a role in the survival of clusters, in the sense that more concentrated polytropes (larger index $n$) resist dissolution better. 
  
\section{Time-evolution and other considerations}\label{sec:discussion} 
The analytic  
approach of Section 3 did not aim to relate final equilibrium to initial conditions, as we have completely  
ignored details of the time-evolution. We turn here to basic considerations 
of cluster dynamics that were left out until now. \newline 

Adams (2000) has recently revisited the problem of forming bound clusters through dynamical evolution. 
His analytical approach 
 leads to the conclusion that in order to form bound clusters, of low  \sfe\ averaged over the 
entire cluster, the \sfe\ must peak with the central  stellar  
density. In his problem, therefore, the large concentration of stars results from efficient star formation 
in a small volume. Our own results would support this view, in as much as we find 
 the more centrally concentrated polytropes to dissolves at lower \sfe. We did not, however, consider an  
 \sfe\ varying 
radially with the density.  The problem would seemingly be solved from the freedom in 
 profiling the \sfe\ with  the stellar potential. Tenorio-Tagle et al. (1986) had 
 noted that gas density gradients have a strong impact on the outcome, since more bound 
gas (steep gradients) requires more mechanical energy for removal, which demands more time 
 under a  constant UV flux: this in effect reduces the ratio of dynamical time to gas-evacuation time
   and shifts the dynamics toward a slow evolution regime, which we briefly outline below. \newline 
 
  In condensed star clusters, the dynamical time $\tcr  $ at the half-mass radius may be long compared to a finite 
 gas-evacuation timescale $\equiv\tau$, while being shorter than $\tau$ at the centre of the cluster. When  
$\tcr/\tau \ll 1$, the integral over an orbital path  $l$ 
\begin{equation} J_\phi \equiv  \oint_l \boldv{v} \rmd \boldv{x}  \propto  v\, r \propto (E/M)^{1/2} \, r \propto M^{1/2}\, r^{1/2} \label{eq:adiabat} \end{equation}
is an adiabat so $J_\phi = $ constant during evolution (Weinberg 1993; BT+87, \S 3.6). 
 Here $M$ is the cluster mass, $E$ the binding energy and $v,r$ the orbital velocity and 
radius of a star, respectively. 
 The regime $\tcr/\tau \ll 1$ reduces to an adiabatic, quasi-static transformation of 
  the initial mass profile, with mass loss acting as a  positive source of energy. Notice that in order for 
(\ref{eq:adiabat}) to apply, no limit 
is set on the fraction of mass lost through gas expulsion, only that it proceeds slowly: then no stars are lost 
through adiabatic evolution. Indeed under such condition clusters prove much more robust to total disruption 
(Lada et al. 1984; Goodwin 1997). 
 From (\ref{eq:adiabat}) we deduce $\tcr \propto \sqrt{r^3/M} \propto r^2/J_\phi$. Since 
the system expands, it will move out of the adiabatic regime $\tcr/\tau \ll 1$ (i.e., $\tcr \approx \tau$) 
if the expansion factor 

\begin{equation} \frac{ R_{final} }{ R_{initial} } = \sqrt{ \frac{\tau}{t_{{\rm cr},initial}} } \ . 
\label{eq:expansion} \end{equation}
We can draw constraints from observations of Scorpius clusters 
 which suggest an expansion factor of up to five between clusters of different ages (e.g. Brown 2001). 
If we take $\tau \approx 10^5$ years as reference (see the introduction), and inserting a ratio = 5 on the left-hand side of (\ref{eq:expansion}), the central 
region of a cluster would evolve adiabatically throughout the gas expulsion phase 
if the core $\tcr$ is initially on the order 
of $ 4 \times 10^3 $ years. For a condensed cluster such that  initially $\tcr \gg \tau$ at the half-mass radius,  
we may assume that $\tcr \propto 1/\sqrt{G\langle\rho\rangle} \gtabout \tau $  from that point on. 
The ratio of central to half-mass densities  
$\rho(0)/\rho_{1/2} = \tcr^2(r_{1/2})/\tcr^2(0) > \tau^2/\tcr^2(0)$ would then be $> (10^5/4\times 10^3)^2 = 625 \sim 10^3$. 
 A Michie-King cluster model (cf. King 1966) with $\Psi/\sigma^2 \gtabout 9$ meets this condition, however a 
Plummer sphere does not. 
This points to yet greater dependence on the \df\ than was obtained for the family of polytropes. 
 Results obtained for the King family of \df\ will be presented in Paper II.


  This analysis reinforces our view that instantaneous gas removal provides the most severe criterion 
for the survival of bound cluster, since any realistic situations will be a compromise between rapid 
gas removal  and  the adiabatic  ($\tcr \ll 
\tau$) regime. 
Hills (1980) had pointed out that explosive mass loss may yet occur at a time when the cluster has not 
fully reached dynamical equilibrium. If the stars' velocities were small initially, the cluster would 
contract radially  until $T/|W| \approx 1$ so that $ Q = 2 T / |W| = 2$ at most (e.g., Boily, Athanassoula \& Kroupa 2002).  Two  possibilities 
may occur: in the first one, the gas is drawn in with the stars so that the mass ratio of gas to stars 
is unchanged along 
the path of the stars. In this case we  may compute the effect of instantaneous gas-expulsion
 by removing stars with positive energy and setting $Q > 1$ in the \df\ (see Appendix B).  
We find for Plummer spheres 
  that total disruption would occur if the \sfe\ were less than about 48.8\% when the virial ratio assumes its 
maximum value of $Q = 2$.  
Clearly this has the contrary effect to the one 
sought. A second possibility is that the gas is expelled when the stars have not 
acquired large velocities and $Q < 1$, a more likely situation 
 given that a cluster initially at rest would spend more time in 
this dynamical phase as can be deduced from  Kepler's third law. 
We find in this case for a Plummer sphere and for $Q \approx 1/4$ 
 bound stars at an \sfe\ $\approx 24\%$, as in observations of 
star forming regions. This provides a solution for bound clusters where stellar ages are 
less than a mean dynamical  cluster crossing time and virial equilibrium has not yet been established. 
 This view has to be weighted against the likelihood of observing clusters out of equilibrium and the reduced time 
available to form stars. 


\section{General conclusions} 
 We have presented a simple method for  determining the fraction of stars that will 
survive a phase of rapid gas loss following the birth of the first heavy stars to form a bound cluster. 
 This approach, based on an iterative selection of stars by binding energy, is a modified version 
of the problem discussed by Hills (1980). 
 The star formation efficiency $\epsilon$ is taken as free parameter which is held constant throughout the 
cluster. Some  highlights from this study are: \newline 

1) The choice of distribution function determines the  fraction of bound stars that remain to form a bound cluster 
following the rapid loss of gas  (cf. Eq.~\ref{eq:escape}). 
Taking account of the \df\ allowed us to estimate more accurately the 
bound mass than the predicted disruption for \sfe\ $< 50\% $ based on the virial theorem alone 
(cf. Eq.~\ref{eq:ratio}).
 
In Appendix A, we show using a harmonic oscillator model that a large fraction of a
self-gravitating stellar system in virial equilibrium will survive gas
expulsion despite a low \sfe\ ($<0.4$) if the 
stellar velocity distribution function favours stars with low velocities. 
It is, therefore, important to measure stellar  velocities in a 
cluster-forming molecular cloud core to obtain observational constraints
on the shape of the \df.

2) Analytic solutions were obtained for all polytropes of index $n \ge \frac{3}{2}$. 
   Polytropes with index $n \gg 1$ yield a bound fraction for very low \sfe, easily matching the 
  range suggested by observations.  
 
 These findings are backed up by  a series of N-body 
calculations, where the orbital evolution of individual stars is fully taken into account: such calculations 
 provide a good match to the solution curve (\ref{eq:lpara}) for the Plummer sphere (Fig.~2). 
This encourages us to apply (12) to other \df\ and to consider the dynamics fully in order 
 to identify which, if any, initial conditions \df\ may lead to bound 
clusters despite very low \sfe. 
We propose to cover these topics  in greater details  in a companion paper (Paper II). 

\section*{acknowledgments} We are very grateful to D.C. Heggie for a thorough read of a draft version 
of this paper. His insight led to a much-improved section 4. We thank him and A. Lan\c{c}on for 
comments that led to clarifications of the arguments. CMB wishes to thank  
A.G.A. Brown and E.K. Grebel who took him gently through 
the observational literature. We are grateful to them for their help and suggestions. 
 We extend our gratitude to an anonymous referee for constructive comments. 
CMB was funded under the SFB~439 program in Heidelberg  
 and wishes to thank his host R. Spurzem at the ARI for unwavering support.

\section*{Appendix A: Bound orbits in a harmonic potential} 
To see how stars on different orbits may escape at different rates, as argued in Section~2,  
consider a self-gravitating  homogeneous mass distribution of density $\rho$. 
 The stellar orbital time $\propto \tcr$ is then independent of position. 
 	Stars are often distributed uniformly in space over a small volume 
so this situation has bearing on actual systems. 
 Our aim is to relate the membership of a star to its orbital parameters  prior to gas expulsion. 

 The equation of motion for this case 
reduces to that of a harmonic oscillator which in one dimension reads  

\begin{equation} \ddot{r}(t) = \nabla\phi(r) = \frac{GM(<r)}{r^2} = - \omega^2\, r(t) \label{eq:harmonic} \end{equation} 
where we set $\omega^2 = 4\pi G\rho/3 $. Integrating (\ref{eq:harmonic}) we have 

\begin{equation} r(t) = r_o \, \cos( \omega\, t + \theta_o ) \label{eq:harmonic.solution} 
\end{equation} 
where the amplitude $r_o$ and angle $\theta_o$ are set by the initial conditions. 
 When the homogeneous region is cut off at radius $r = R$, the potential everywhere in space is 

\begin{equation} \phi(r) = - \frac{3}{2} \frac{GM}{R} + \frac{\omega^2}{2} \, r^2 
\ \ r \le R \ , \label{eq:harmonic_potential} \end{equation}
and $\phi(r) = - GM/r $ for $r > R$. 
The derivative of $\phi$ is discontinuous at $ r = R$ but this has no bearing on the 
argument presented here since we consider motion with $r(t) < R$ only.  We imagine an 
ensemble of stars sharing the same orbit of amplitude $r_o$ 
 but otherwise phase-mixed. The energy per unit 
mass for that orbit 

\[ {\cal E} = \phi(r) + \half\, v^2 (r) = \phi(r_o) \] 
is modified at time $t = t_o$ when a fraction of the mass is lost and only stars remain: 
$\phi_\ast (r) = \epsilon \phi(r) $ is now the gravitational potential. The new energy 
 per unit mass at radius $r$ 

\[ {\cal E}_\ast = \phi_\ast(r) + \half\, v^2 (r) = \epsilon \phi(r) + {\cal E} - \phi(r) \] 
is  positive if $v^2 > -2 \phi_\ast(r) = -2\epsilon\phi(r)$ which 
 is possible provided (regrouping terms and remembering [\ref{eq:harmonic_potential}] ) 

\begin{equation} 
\omega^2 r_o^2 + \epsilon\ \omega^2 r^2 \ge \epsilon\ \frac{3\,GM}{R} = \epsilon\ 3\, \omega^2R^2\ . \label{eq:harmonic_unbound} \end{equation}
Since the original orbit (\ref{eq:harmonic.solution}) has $r(t) \le r_o$, we find from (\ref{eq:harmonic_unbound}) a 
minimal amplitude $r_o$ for unbound orbits as function of the \sfe\ $\epsilon$ : 

\begin{equation} \sqrt{\frac{3\,\epsilon}{1+\epsilon} }\ \le \ \frac{r_o}{R} \ < \ 1\ ,
\label{eq:harmonic_ro} \end{equation}
where the upper limit follows from our definition of $R$. 
Equation  (\ref{eq:harmonic_ro}) states that all stars on 
orbits with amplitude $r_o$ sufficiently 
small  will remain bound, while at least a fraction of those on large-amplitude 
orbits will escape. Otherwise said,  stars on low-binding energy orbits, ie which achieve high velocities 
at some point on their orbit, are more likely to unbind following gas expulsion. 
%
 All orbits with 
 $r_o \ll R$ would  lose none or a smaller fraction of stars than those that visit the edge of 
the system. 

\section*{Appendix B: Deviations from dynamical equilibrium} 
The virial ratio $Q \equiv 2 T / |W|$ = 1 for a system in dynamical equilibrium. In practice the \df\ may 
not satisfy this condition exactly and this requires modifications to (\ref{eq:plummer_lambda}). We consider the 
case of the Plummer sphere only but a similar treatment would apply to all polytropes.  \newline 

Substituting absolute values for $E$ in (\ref{eq:plummer}) allows to treat sub- as well as super-virial dynamics. Then 
the integrals (\ref{eq:plummer_lambda}) read 

\begin{equation} 
1 - \lambda_e = \frac{\displaystyle \int_{\epsilon^{1/2}}^{Q^{1/2}} ( \arrowvert 1 - u^2\arrowvert 
)^{7/2} \, u^2\,{\rm d}u }{ \displaystyle \int_{0}^{Q^{1/2}}( \arrowvert 1 - u^2\arrowvert 
)^{7/2} \, u^2\, {\rm d}u} = 1 - \frac{{\cal F}(\epsilon)}{{\cal F}(Q) + {\cal G}(Q)}  \label{eq:plummer_lambdaQ}
\end{equation} 
where the function ${\cal F}$ is defined by 

\begin{equation} {\cal F}(x) \equiv \left( { \sin^{-1} 
x^{1/2} - p(x)\,x^{1/2}\sqrt{1-x} }{}\right) \end{equation} 
with $p(x)$ defined by  (\ref{eq:plummer_lambda}), and 

\begin{equation} {\cal G}(Q>1) \equiv \frac{7}{256} \left[ \sqrt{Q}\sqrt{Q-1}\, p(Q) + \log( \sqrt{Q} + \sqrt{Q-1} )\right]
\end{equation} 
and ${\cal G}(Q\le 1) = 0.$ Clearly the numerator in (\ref{eq:plummer_lambdaQ}) 
evaluates to zero if $\epsilon \ge Q$ when the solution $\lambda_e = 1$. Since the \sfe\ $\epsilon$ is bounded 
to $[0,1]$, this is possible only for sub-virial conditions $(Q<1)$. We recover (\ref{eq:plummer_lambda}) by setting 
$Q = 1$ in (\ref{eq:plummer_lambdaQ}) since ${\cal F}(1) = \pi/2$. \newline 

Sub-virial conditions are relevant to the problem of forming bound clusters (e.g. Klessen \& Burkert 2000; Clarke et al. 2000), 
with the caveat  that sub-virial conditions at the gas-expulsion time require synchronous star formation throughout the cluster to 
within a fraction of the dynamical time, when the spread in colours suggest at least a formation time of a few million years 
(Hartmann et al. 2001). 
 We wish to show that a point of the parameterised solution 
 (\ref{eq:lpara}) and (\ref{eq:epara}) shifts to lower $\epsilon$ as $Q$ is reduced from 1. 
 Treating $Q$ as the variable in (\ref{eq:plummer_lambdaQ}) and differencing we have 

\begin{equation} \frac{\rmd\, \epsilon(x_o)}{\rmd Q} = \frac{x_o}{{\cal F}(x_o)} \, {\cal F}^\prime (Q) \ ; \  \frac{\rmd\, \lambda_e(x_o)}{\rmd Q} = - \frac{\lambda_e(x_o)}{{\cal F}^2(x_o)} \, {\cal F}^\prime (Q)\ , 
\end{equation} 
where $x_o$, the parameter defined in (\ref{eq:defx}), identifies a point of the solution curve $(\lambda_e,\epsilon)$ when  
$Q = 1$. The derivative 

\[ {\cal F}^\prime (Q) = \frac{128}{7} \, \sqrt{Q} \, \left( 1 - Q \right)^{7/2} > 0 \,\ \ \forall \ \ Q < 1\ . \] 
It is easy to see from the above equation and (32) that reducing the virial ratio $(\rmd Q < 0) $ shifts the abscissa $\epsilon$ 
to smaller values while increasing the bound fraction $\lambda_e$, for any point $x_o$. 
Therefore `cooler' initial conditions 
lead to bound clusters for lower \sfe\ $\epsilon$. The gap from finite $\lambda_e$ to zero (total dissolution) is 
 more pronounced but occurs at lower values of $\epsilon$ as $Q$ is reduced. 

Since ${\cal F}^\prime(Q)$ is near zero around $Q = 1$, the 
net effect of reducing the virial ratio $Q$ from unity is at first marginal and only becomes significant once 
$Q \lesssim \half$. Similar conclusions apply to super-virial ($Q>1$) conditions. Total dissolution  occurs for 
larger \sfe\ $\epsilon$ in these cases. A cluster collapsing from rest would reach $Q = 2$ at most during violent 
relaxation.  Dissolution occurs for an $\epsilon \approx 48.4\%$, only marginally larger than the 44.2\% value obtained for
equilibrium configurations. The solution for $Q=2$ and $\epsilon = 1$ (\sfe\ of 100\%) 
yields a bound fraction $\lambda_e \approx 0.91$, 
such that 9\% of the stars escape. This is remarkably similar to what is observed in 
N-body realisations of collapsing cold spheres. 
 
\flushbottom  
\newpage 


\begin{table*}
\caption{Critical \sfe\ $\epsilon$ and bound fraction $\lambda_e$ in parameter form (\ref{eq:polytrope}) for polytropes of index $n$.}
\label{tab:polytropes}
\begin{center} 
\begin{tabular}{ccccl}
$n$ &  $x$ & \sfe\ $\epsilon$ & $\lambda_e$ &  Comments  \\\hline  
    &      &                  &             &            \\  
 $\frac{3}{2}$  &1.000 & 1.000  &  1.000      &            \\
 2  &  0.528 & 0.857          &  0.616      &            \\
 3  &  0.470 & 0.698          &  0.673      &             \\
 4  &  0.296 & 0.543          &  0.546      &  \sfe\ close to (\ref{eq:ratio})  \\ 
 5  &  0.225 & 0.442          &  0.509      &  Plummer sphere \\
 6  &  0.182 & 0.373          &  0.485      &               \\
 8  &  0.131 & 0.283          &  0.463      &                \\
10  &  0.102 & 0.228          &  0.448      &                \\ 
20  &  0.044 & 0.115          &  0.424      &                 \\
40  &  0.024 & 0.058          &  0.412      &                 \\
$\infty$ & 0.000 & 0.000      &  -----      & Isothermal sphere \\ 
\end{tabular}
\end{center} 
\end{table*}


\setlength{\unitlength}{1in} 
\begin{figure*} 
\begin{picture}(4.5,5.)(0,0) 
        \put(.0,.00){\epsfxsize=0.7\textwidth\epsfysize=0.7\textheight
\epsfbox{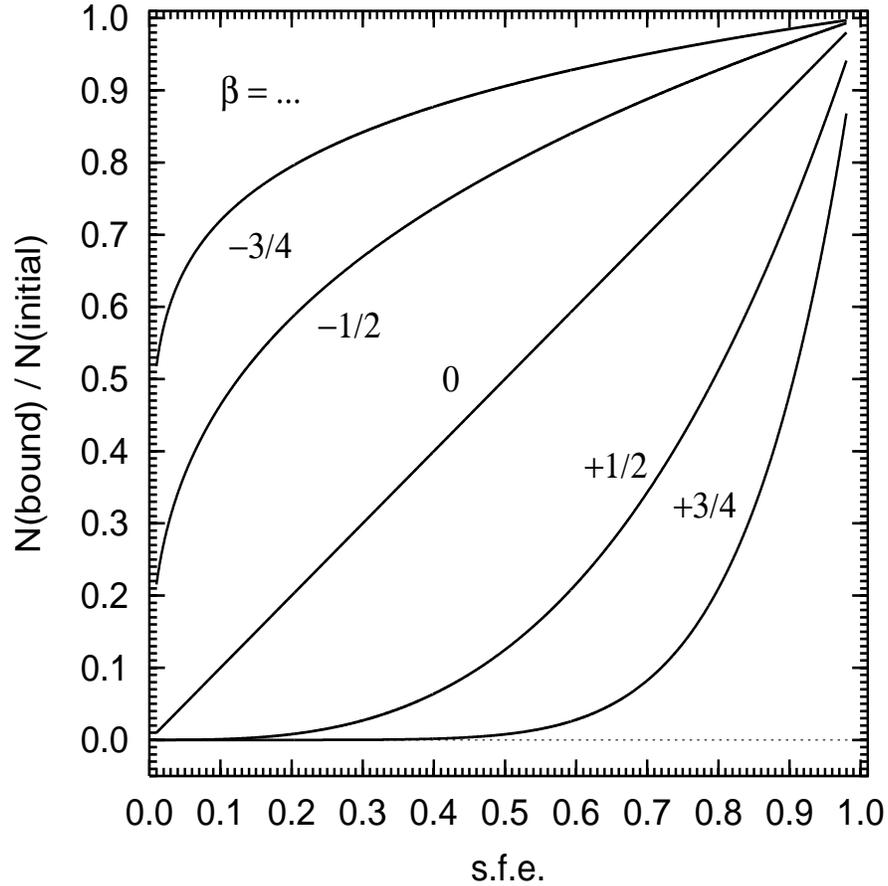}
                  }
\end{picture}  
\caption{Ratio $M^b_\star/M_\star$
 of  bound  
 to initial stellar masses  as function of  star formation efficiency for power-law 
\df\ (\ref{eq:powerlaw}). Five values of the power index $\beta$ are displayed. Those with $\beta > 0$ are weighed in favour of high-velocity stars: 
the fraction of bound stars drop more sharply with decreasing \sfe.}
\label{fig:powerlaw}
\end{figure*} 

\newpage 
 
\setlength{\unitlength}{1in} 
\begin{figure*} 
\begin{picture}(4.5,4.5)(0,0) 
        \put(0.25,0.1){\epsfxsize=0.6\textwidth\epsfysize=0.5\textheight 
\epsfbox{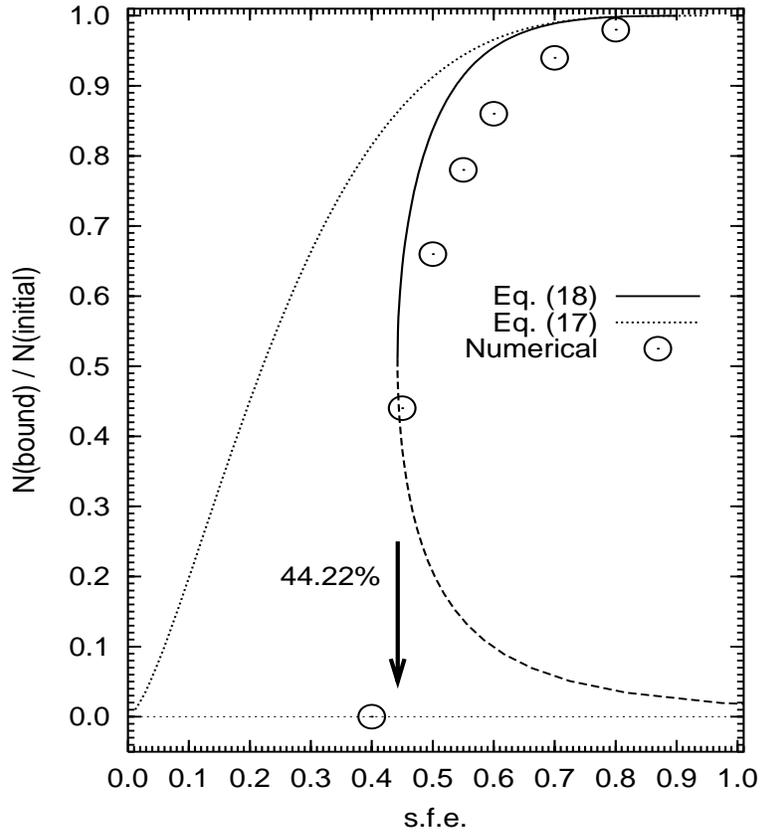}
                  } 
\end{picture}  
\caption{Ratio $\lambda_e = M^b_\star/M_\star$ of  bound  
 to initial stars as function of  star formation efficiency, $\epsilon$. 
The results for Plummer models derived from the iterative scheme 
(\ref{eq:lpara}) are shown  as the solid line. By contrast, the dotted curve is obtained from 
 (\ref{eq:plummer_lambda}), when no iterations are performed:  in this case 
a finite bound fraction would be expected for any positive \sfe. The results 
of numerical N-body computations are shown as open circles. } 
\label{fig:plummer} 
\end{figure*}

\end{document}